\documentclass[showpacs,aps,twocolumn,showpacs,floatfix,nofootinbib]{revtex4}
\usepackage[dvips]{graphicx}
\usepackage{dcolumn}
\DeclareGraphicsExtensions{.ps,.esp}
\newcommand{\helium}{{${}^4$He}}
\newcommand{\beryllium}{{${}^8$Be}}
\newcommand{\carbon}{{${}^{12}$C}}
\newcommand{\oxygen}{{${}^{16}$O}}
\newcommand{\neon}{{${}^{20}$Ne}}
\newcommand{\magnesium}{{${}^{24}$Mg}}
\newcommand{\silicon}{{${}^{28}$Si}}
\newcommand{\sulphur}{{${}^{32}$S}}
\newcommand{\argon}{{${}^{36}$Ar}}
\newcommand{\calcium}{{${}^{40}$Ca}}
\begin{document}
\title{Alpha Cluster Structure and Exotic States in a Self-Consistent
Model for Light Nuclei}
\author{J. A. Maruhn}
\affiliation{Institut f\"ur Theoretische Physik, Universit\"at Frankfurt, Max-von-Laue-Str. 1,
60054 Frankfurt am Main, Germany}
\author{Masaaki Kimura}
\affiliation{Yukawa Institute for Theoretical Physics, Kyoto 606-8052, Japan}
\author{S. Schramm}
\affiliation{Institut f\"ur Theoretische Physik and Center for Scientific Computing, Universit\"at Frankfurt, Max-von-Laue-Str. 1,
60054 Frankfurt am Main, Germany}
\author{P.-G. Reinhard}
\affiliation{Institut f\"ur Theoretische Physik II, Universit\"at
  Erlangen-N\"urnberg, Staudtstrasse 7, 91058 Erlangen, Germany}
\author{H. Horiuchi}
\affiliation{Physics Department, Kyoto University, Kyoto 606-8052, Japan}
\altaffiliation[Present address: ]{Research Center for Nuclear Physics, Osaka University,
Mihogaoka 10-1, Ibaraki, Osaka 567-0047, Japan}
\author{A. Tohsaki}
\affiliation{Suzuki Corporation, 46-22 Kamishima, Kadoma 5710071, Japan}
\date{}
\begin{abstract}
In this paper we examine to what extent traces of $\alpha$ clustering can be found in
mean-field ground states of $n\alpha$-nuclei from \beryllium\ through \argon\ as well as in some
superdeformed states in \sulphur, \argon, and \calcium. For this purpose we
calculate the overlap of the mean-field Slater determinant with one containg pure Gaussians and
perfect spin and isospin symmetry, optimizing the overlap by varying the $\alpha$-particle positions
and radii. In some cases a coherent sum over different configurations is also employed. We find
quite large overlaps for some of the lighter systems that diminish for nuclei above \neon\ but
again strong clustering in \argon.
\end{abstract}
\maketitle
\section{Introduction}
For light $n\alpha$ nuclei, i.~e., nuclei composed of $2n$ protons and $2n$ neutrons, a description
in terms of $\alpha$-clusters has enjoyed considerable success over many years. In their simplest
form, these correspond to Mean--Field with a restricted wave function made up of
$\alpha$-clusters: quadruples of particles with both spin and isospin orientations and with Gaussian
wave functions in space centered at given positions. These positions are then varied to obtain an
optimal many-body wave function. Later variations of this method also allow variation of the radii,
deformed clusters, and exploit the computational efficiency of the approach to calculate projected
states (see \cite{project} and references therein).

In the full Hartree--Fock or Mean--Field approach, on the other hand, the wave functions can be quite complicated
in their spatial behavior and indeed in many modern calculations are freely variable over a
coordinate grid, so that typically thousands of values are varied to optimize the many-body wave
function. Since the Mean--Field state is a pure Slater determinant, it does not contain
correlations in the conventional sense, which seems to exclude correlated objects like
$\alpha$-particles. A closer examination reveals, however, that correlations come in through the
mean field. The identical wave functions assumed for each set of four particles in an
$\alpha$-cluster model are present in the full Mean--Field solution to a certain extent, because
the eigenfunctions in the same mean field potential will be filled with four particles, the main
cause for deviations from the identical wave functions being spin-orbit coupling (destroying spin
degeneracy) and Coulomb effects (destroying isospin degeneracy).

In this paper we address the following questions:
\begin{enumerate}
\item To what extent do the Mean--Field states agree with an
$n\alpha$-cluster model, i.~e. can they be represented by cluster model
wave functions?
\item Does this agreement depend in a systematic way on spin orbit
strength and the mass of the nuclei?
\item Is the Mean--Field state better represented by a correlated
cluster state, in the sense of a coherent sum over different cluster
configurations?
\item Are there indications for more exotic cluster states producible
in the Mean--Field approach?
\end{enumerate}

It is, of course, interesting to compare the results with cluster-model interpretations of these
nuclei. In the following, we will refer to the geometric configurations suggested by Zhang, Rae, and
Merchant \cite{zhang} based on a cranked version of the Block-Brink $\alpha$-cluster model. In some
cases also a comparison to deformed-basis antisymmetrized molecular dynamics (AMD) \cite{defamd}
calculations was performed.

\section{The Mean--Field calculations}
\subsection{Method of Calculation}
We represent the single-particle wave functions on a Cartesian grid with a grid spacing of
1~fm. The grid size is typically $24^3$ for ground states and $36\times24^2$ for superdeformed
states. This accuracy was seen to be sufficient to provide converged configurations.  The numerical
procedure is the damped-gradient iteration method \cite{gradient}, and all derivatives are
calculated using the Fourier transform method.

In the case of shape isomeric states, convergence is sometimes difficult to establish. Experience
has shown that the observation of the change in total energy alone is not sufficient to judge
convergence: this value usually typically decreases rapidly to $\delta E/E\approx10^{-9}$, which may
indicate simply that the calculation is getting stuck and convergence is really stalled. A much
better criterion turned out to be the mean-square deviations in the single-particle energies, summed
over all states. These uncertainties thus measure how far the states still are from true eigenstates
of the single-particle Hamiltonian.

In practice it was often seen that the total energy seemed to converge perfectly well, while the
fluctuations in the Hamiltonian stay at a relatively high level of $10^{-3}$. In such cases usually
the system will stay in this configuration for several thousand iterations and then transit to the
ground state relatively rapidly. The underlying situation can be visualized as a steep descent which
arrives at some saddle point, where the direction of the descent must be modified strongly, keeping
the system at this point for a very large number of iterations. Alternatively it may be caught in a
shallow local minimum from which the accumulation of small changes in the wave functions enable the
system to escape during further iterations.

\subsection{Interactions and symmetries}
We take four different Skyrme forces which all perform very well concerning nuclear bulk properties
but differ in details: SkM$^*$ as a widely used traditional standard \cite{Bar82}, Sly6 as a recent
fit which includes information on isotopic trends and neutron matter \cite{Cha97a}, and SkI3 as well
as SkI4 as recent fits which map the relativistic iso-vector structure of the spin-orbit force
\cite{Rei95a}.  SkI3 contains a fixed isovector part analogous to the relativistic mean-field model,
whereas SkI4 is adjusted allowing free variation of the isovector spin-orbit term. Thus all forces
differ somewhat in their actual shell structure. Besides the effective mass, the bulk parameters
(equilibrium energy and density, incompressibility, symmetry energy) are comparable.
In addition we consider two relativistic mean-field models. $NL3$ is a commonly used parameter set
that includes non-linear interactions of the $\sigma$ mesons that generate the scalar interaction between the
nucleons \cite{ring}. The model $\chi_m$ is based on a chiral flavor-SU(3) effective Lagrangian,
which is discussed in detail in \cite{schramm}.

It will be seen that the results are generally in quite good
agreement, with mainly the spin-orbit strength producing differences among the Skyrme forces;
SkM* systematically deviates somewhat from the others. For NL3 and $\chi_m$ the spin-orbit interactions
are a direct consequence of the relativistic description of the nucleon wavefunctions. The treatment of the center-of-mass
motion is considerably different in SkM* from the other forces; since we do not include a
center-of-mass correction, there are considerable differences in the binding energies of light nuclei.

The wave functions were not subject to any symmetry requirements, having neither good spin nor good
parity quantum numbers; in fact, the initialization purposely avoided symmetries to make sure that
these were not accidentally maintained.

The calculations included pairing in the BCS approximation with a delta force, taking into account
three times the number of occupied states without pairing. The parameters were the ones usually
associated with each Skyrme force.

\subsection{Observables}
The principal observables that will be quoted are the total binding or excitation energy, the
spin-orbit energy, the deformation, and the intrinsic quadrupole moment.  The quadrupole moment was
calculated in the standard way as
\begin{equation}
Q=\int\,{\text d}^3r\rho(\vec r)(3z^2-r^2)
\end{equation}
for the case that the $z$-axis is the axis of axial symmetry. A deformation
parameter is then obtained as
\begin{equation}
  \beta_{20}=\sqrt{\pi\over5}\frac{Q}{AR^2},
\end{equation}
$A$ being the mass number and $R$ the mean-quare radius of the system. This prescription is, of
course, not unique, since the dependence on the mean-square radius is somewhat arbitrary, but
follows \cite{nazar1}. Oblate deformations are recognizable by the negative sign in $Q_{20}$ and
$\beta_{20}$.

\section{Methods of Analysis}
To investigate possible molecular structure, three different methods of analyzing the mean-field
Slater determinant $|\Psi\rangle$ were employed. In this section the procedures are described, while
detailed results for the various nuclei will be given in Section~\ref{gsnuclei} and~\ref{hyperdef}.

\subsection{$n\alpha$-cluster model}
We investigated the clustering structure by constructing a model many-body wave function with
clustering aspects built in and maximizing its overlap with the Slater determinant obtained from the
Mean--Field calculation. As the single-particle wave functions have no good quantum numbers aside
from the energy, only the overlap of the complete Slater determinants can be meaningfully
investigated. In case of the relavistic models we used the upper components of the wavefunctions to calculate the
overlap.

In the simplest case, which we shall refer to as the $n\alpha$-model, the model wave function is
constructed completely out of $\alpha$-particle configurations.For a nucleus with $A$ nucleons, we
use $N_\alpha=A/4$ clusters located at $\vec r_k$, $k=1,\ldots N_\alpha$ and with radius parameters
$\sigma_k$.  Depending on the total number of $\alpha$-particles, we usually constrained their radii
to be equal, but sometimes, when total computing time allowed, also allowed them to vary to see
whether the overlap is thereby improved. The resulting improvement was of the order of a few percent
typically, so that in most cases we will quote results only with one common value of $\sigma$
adjusted for maximum total overlap.

The spatial part of the wave functions then is given by
\begin{equation}
\phi_k(\vec r)=A_k\exp\left[
-\frac{(\vec r-\vec r_k)^2}{2\sigma_k^2}\right],\quad k=1\ldots N_\alpha.
\end{equation}
with $A_k$ a normalization factor that is computed numerically. The numerical normalization factor
calculated on the same spatial grid with 1~fm grid spacing as the Mean--Field wave functions was
found to agree with the analytic one to five significant digits, so that the spatial grid
appears sufficiently accurate for the overlap calculations. Each of these wave functions occurs four
times in the model Slater determinant: for proton and neutron and with spin up or down.

Note that the $n\alpha$-cluster single-particle wave functions are not orthogonal. As a consequence,
the norm of the model Slater determinant must be calculated as the determinant of the
single-particle overlaps
\begin{equation}
  {\cal N}=\det \left(\langle \phi_j^* |\phi_k\rangle\right).
\end{equation}
This determinant can become very small if the wave functions have large overlaps, which occur
frequently in the optimal configurations, indicating that the cluster description does not
correspond to clearly separated $\alpha$-particles, but generates the Mean--Field states largely
by antisymmetrization. For example, two Gaussians with the centers placed very close to each other
generate an odd-parity state through antisymmetrization.

The overlap ${\cal O}=|\langle\Psi|\Phi\rangle|^2$ given in the tables below is always the absolute
square of the matrix element between the Mean--Field Slater determinant and the normalized
$n\alpha$-cluster wave function $|\Phi\rangle$, so it describes the probability for the Mean--Field
state to agree with the cluster model state.  Since the Mean--Field wave functions have good
isospin but are mixed in spin projection, the total overlap could be simplified by decomposition
into a proton and a neutron factor. Although the effects of Coulomb are practically negligible, the
calculation did treat the two isospins separately.

\subsection{Calculation of $n\alpha$-cluster configurations}
The most effective method to search for the optimal positions and sizes of the $n\alpha$-cluster
positions was a search with an optimization algorithm starting from random positions (within a box
roughly corresponding to the nuclear volume) and a standard starting value for the radius parameters
of 1.8~fm. We found that optimizing the logarithm of the overlap of the two Slater determinants with
the algorithm of Stewart \cite{stewart} (available as routine ``dmnfb'' in NETLIB) led to very rapid
convergence even when the individual particles were placed very badly initially with a starting overlap of
$<10^{-15}$ with the Mean--Field Slater determinant.

The final configurations could in all cases be decomposed into
\begin{enumerate}
\item isolated $\alpha$-particles in a very well determined position
  characterizing the cluster geometry, and
\item groups of $\alpha$ particles spaced very closely at distances
  typically $<0.1$~fm. These represent cores made up of larger nuclei, usually \carbon\ or
  \oxygen. Since the wave functions in these clusters serve mostly to generate higher shell-model
  states through antisymmetrization, their geometrical layout appears random. It was found that the
  overlap with the mean-field wave functions changes only by about 2\% when the distance between the
  particles increases from 0.01~fm to 0.8~fm. It is largest for the smallest distance, but this can
  cause numerical problems because of the nearly degenerate wave functions.
\end{enumerate}

It should be noted that the positions found usually showed some alignment with the Cartesian axes
even for spherically symmetric nuclei, since the representation on a Cartesian grid violates that
symmetry to a small but noticeable degree.

In some cases a fixed geometry was also employed to check how different cluster geometries compete
in describing the HF ground state. In these cases a small number of geometric quantities was varied
to achieve optimal overlap.

\subsection{Model of core nucleus with additional $\alpha$-particles}
In some cases, a model wave function was used where a core of \carbon\ or \oxygen\ replaced the
corresponding number of $\alpha$-particles. Note that the wave functions used for these were simply
static Mean--Field solutions for those core nuclei, without any adjustable parameters (except for
the position of their centers).  In the cases where three or four $\alpha$-particles tended to
contract into the same location, this may produce more stable numerical results because of the norm
going to zero for the $\alpha$-particle case. The most interesting application of this technique is
to the interpretation of the strongly deformed state of \sulphur\ as an \oxygen+\oxygen\
configuration.

\subsection{Overlap with a collective cluster space}\label{gcmmodel}
In some cases it was tried to allow some parameter determining the $n\alpha$-cluster configuration
to vary in the model wave function and to calculate the overlap with the collective space generated
in this way. As an example, in the case of the strongly deformed \sulphur\ with Slater determinant
state $|\Psi\rangle$ interpreted as an
\oxygen+\oxygen\ configuration, the Slater determinants $|\Phi_i\rangle$ consisting of the
antisymmetrized combination of two \oxygen\ nuclei at distances $d_i$, $i=1\ldots N$ were used as a
nonorthogonal basis. The matrix elements of the overlap kernel for this basis are
$A_{ij}=\langle\Phi_i|\Phi_j\rangle$. The probability for the state $|\Psi\rangle$ to be
within this space is given by
\begin{equation}\label{eq:gcmover}
{\cal O}=\sum_{i,j=1}^N \left(A^{-1}\right)_{ij}
\langle\Phi_j|\Psi\rangle\langle\Psi|\Phi_i\rangle.
\end{equation}
The result depends on $N$, but converges rapidly as soon as the collective space is sufficiently
spanned. The value of $N$ of course needs to be chosen carefully. We found that numerically the
results remain stable for $N<20$, but usually $n\approx5$ is already sufficient for convergence.

The criterion of Eq.~(\ref{eq:gcmover}) to extend the Slater stgate $|\Psi\rangle$
can be reproduced by a coherent superposition of $\alpha$-model states
\begin{equation}
 |\Psi^{\rm(model)}\rangle  = \sum_i|\Phi_i\rangle c_i
\label{eq:GCMalpha}
\end{equation}
with optimized coefficients $c_i$. This ansatz is the discrete version of the celebrated Generator
Coordinate Method (GCM) \cite{gcm} which had, in fact, one of its early successful applications in
the realm of the $\alpha$-cluster model \cite{weiguny}.  Note that the criterion (\ref{eq:gcmover})
compares somewhat different concepts.  The Mean-Field state $|\Psi\rangle$ is the optimum one can
achieve in the space of pure Slater states. The GCM ansatz (\ref{eq:GCMalpha}), on the other hand,
is restricted to Slater states within the $\alpha$-model but goes beyond Slater space in that it can
include substantial correlations in its coherent superposition. In spite, or just because, of this
conceputal difference, it is most interesting to investigate the mutual overlap of these two lines
of approach in Eq.~(\ref{eq:gcmover}).

\section{General features of the results}
A general feature of all nuclei considered was that the ground states are axially symmetric as well
as reflection symmetric for all the forces considered, although neither the initialization nor the
computation enforced these symmetries..

We found in all cases that the converged stationary states showed no pairing, so that calculations
{\em without} pairing might appear sufficient.  There is, however, the danger that in iterations
using the restricted single-particle space of the occupied wave functions the two fragments may get
stuck in a local configuration. It is known that without pairing the deformation energy curve is an
envelope of many relatively steep curves corresponding to specific configurations and not all of the
narrow minima produced for such a configuration remain minima in the envelope.

In practice it was found that only part of the highly deformed states were still produced in the
calculation if pairing was included. Although again in the final result pairing disappeared, we
therefore discuss only these ultimately stable configurations.

The freedom allowed by the three-dimensional code was found not to be necessary in these pure
Mean--Field cases, as
the stationary states in all cases turned out to axially symmetric. A comparison with
$n\alpha$-cluster model wave functions of course required three-dimensional geometry.  A code with
axial symmetry and the facility for quadrupole constraints was used in some cases to support the
results and check deformation dependence.

\section{Ground states of $n\alpha$ nuclei}\label{gsnuclei}
In the following we summarize the properties of the $n\alpha$-nuclei below \calcium. Density contour
plots are shown only for cases with interesting structure and only for SkI3; in these the contour
lines are spaced by $0.01\;{\text fm}^{-3}$. The contour lines sometimes appear non-smooth; this is
because the graphics program does not interpolate. The physical wave functions are quite smooth
owing to Fourier expansion.  In all tables, we use $E_{ls}$ for the spin-orbit energy, and ${\cal
O}$ for the overlap with $n\alpha$- or collective cluster configurations.

For some specific nuclei we also mention the overlap between the Mean--Field wave function and the
AMD wave function, and also that between AMD and the $n\alpha$-cluster wave function. Here the AMD
wave function is obtained using the Gogny D1S force \cite{gogny}. The use of a different type of
force will, of course, reduce the agreement with the Mean--Field calculations, but a comparison is
nevertheless instructive.

\subsection{Binding Energies}

\begin{table}
\caption[]{Binding energies in MeV calculated for the nuclei considered in this paper
with the four Skyrme force parametrizations used. \label{tab:EB}}
\begin{tabular}{|c|c|c|c|c|c|}
\hline
 & \helium &\beryllium &\carbon &\oxygen &\neon\\
\hline
Exp.           &28.3&56.5&92.2&127.6&160.6\\
\text{SkI3}    &27.8&49.7&89.3&128.9&156.8\\
\text{SkI4}    &27.7&49.8&89.3&128.6&157.3\\
\text{SLy6}    &27.2&49.0&88.6&127.4&155.9\\
\text{SkM*}    &26.8&50.1&93.5&128.0&157.9\\
\text{NL3}     &33.9&52.9&91.2&128.7&156.6\\
\text{$\chi_m$}&39.3&53.6&88.9&132.3&161.9\\
\hline
\hline
 &\magnesium&\silicon&\sulphur&\argon&\\
\hline
Exp.           &198.3&236.5&271.8&306.7&\\
\text{SkI3}    &194.7&233.0&267.1&304.8&\\
\text{SkI4}    &196.11&234.9&269.4&305.5&\\
\text{SLy6}    &194.4&233.1&268.5&304.0&\\
\text{SkM*}    &197.6&237.9&275.1&305.5&\\
\text{NL3}     &194.1&231.8&265.6&302.2&\\
\text{$\chi_m$}&197.0&235.0&269.5&308.8&\\
\hline
\end{tabular}
\end{table}
In Table~\ref{tab:EB} the binding energies obtained in all of the ground-state calculations are
listed for reference together with the experimental values. All calculations contain a large
contribution from the center-of-mass correction, which is calculated in different ways: the Skyrme
forces SkI3, SkI4, Sly6 and the $\chi_m$-model contain a microscopically calculated correction, NL3
uses the harmonic oscillator approximation, and SkM* a correction to the nucleonic mass
$m\rightarrow m-m/A$.

\subsection{Cluster configuration overview}
\begin{figure}[htb]
\includegraphics{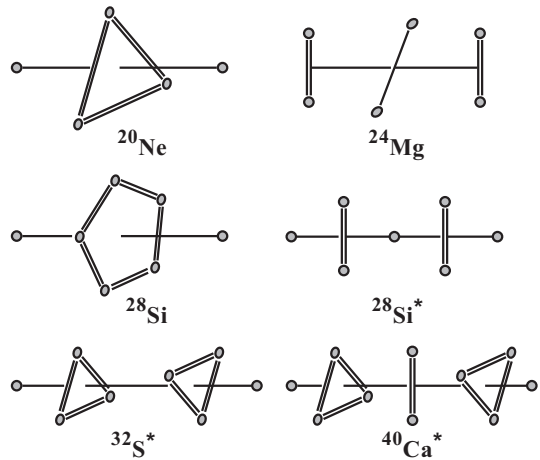}
\caption{Visualization of the cluster geometry present in the Mean--Field ground or excited
superdeformed states (denoted by asterisks) for the
nontrivial cases. The figures only illustrate the topology; no attempt was made to reproduce correct
scale lengths. Double links between particles indicate very small distances. The triangles and
pentagon are perpendicular to the linear links.}\label{fig:structure}
\end{figure}

The optimal cluster configurations obtained are described below in detail. To facilitate
visualization, we show an overview of the geometry in Fig.~\ref{fig:structure}. Those cases where
there is a trivial configuration as well as those without clear structure preference are omitted.

\subsection{The Nucleus \helium}
The $\alpha$-particle itself is of course an extreme case for a mean-field description. We summarize
its properties in Table~\ref{tab:He} mostly because it allows to judge how justified it is to use
Gaussian wave functions for the overlap calculations with an $n\alpha$-cluster structure in heavier
systems and how the optimum radius depends on the force.  For calculating the overlap of the
$\alpha$-particle itself with the Gaussian model wave functions, only the radius parameter $\sigma$
was varied with the optimum value given in the table. In all cases the overlap is remarkably close
to one, showing that the use of the Gaussian is a good approximation. Note that in the case of the relativistic
descriptions the width of the Gaussian is somewhat smaller than for the Skyrme forces, which leads to smaller widths
throughout the $\alpha$-cluster analysis presented in this section.

\begin{table}
\caption[]{Physical properties of \helium\ for the different forces
considered together with data describing the optimum reproduction by
Gaussian wave functions. Note that for the relativistic cases there is no separate $LS$-force, so no $E_{ls}$ values are given.\label{tab:He}}
\begin{tabular}{|c|c|c|c|c|c|}
\hline
Force&$E_{ls}$&$R_{\text rms}$&\multicolumn{2}{c|}{Gaussian Model}\\
&[MeV]&[fm]&${\cal O}$ [\%]&$\sigma$ [fm] \\
\hline
SkI3&0&2.06&99.5\%&1.67\\
SkI4&0&2.07&98.8\%&1.68\\
Sly6&0&2.12&98.8\%&1.72\\
SkM*&0&1.97&99.4\%&1.61\\
NL3&  &1.87&99.4\%&1.54\\
$\chi_m$&&1.89&89.8\%&1.54\\
\hline
\end{tabular}
\end{table}

\subsection{The Nucleus \beryllium}
\begin{figure}[htb]
\includegraphics[width=4cm]{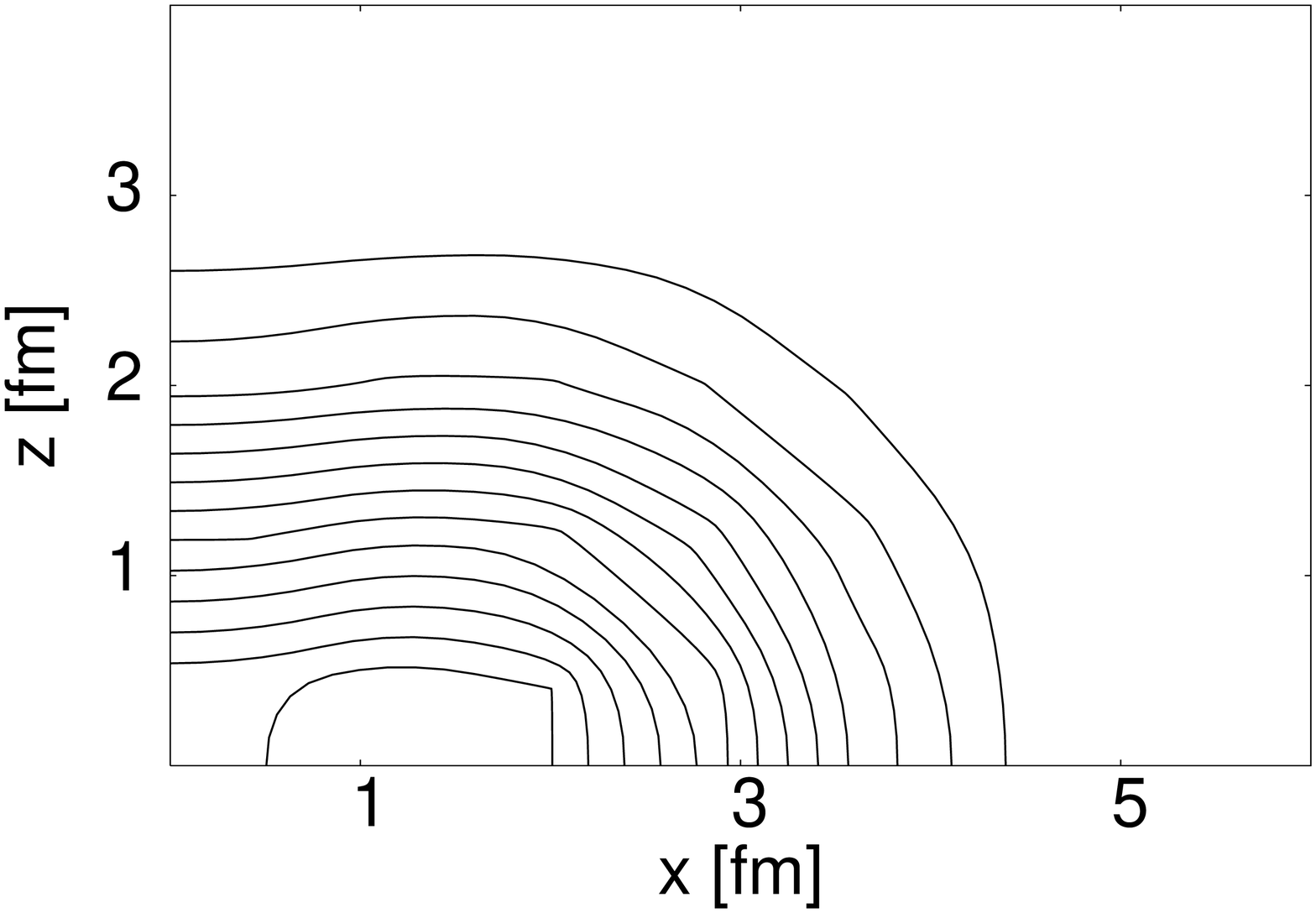}
\caption{Density contour lines for \beryllium.}\label{fig:Be}
\end{figure}

This nucleus is a clear-cut case for a molecular interpretation. For all Skyrme forces considered it
turned out to have a necked-in shape agreeing well with a double-$\alpha$ structure interpretation.
Density isolines illustrate this in Fig.~\ref{fig:Be}, while other properties are summarized in
Table~\ref{tab:Be}. The physical properties are quite similar, with SkM* being the odd force out, as
expected.

The $n\alpha$-cluster analysis naturally found two $\alpha$-particles positioned symmetrically along
the symmetry axis with a distance $d$ and a radius of $\sigma$. The results in the table indicate
that this interpretation works exceedingly well, as even the radius parameters $\sigma$ agree
remarkably well with those for the free $\alpha$-particle. There is, however, a notable
nonorthogonality of the two clusters.
\begin{table}
\caption[]{Physical properties of \beryllium\ for the four Skyrme forces
considered. For the cluster model interpretation the overlap is given together with the optimum
distance and radius of the clusters. The last two columns indicate the overlap with an $\alpha$-GCM
model state and the optimal $\sigma$ obtained. \label{tab:Be}}
\begin{tabular}{|c|c|c|c||c|c|c||c|c|c|}
\hline
Force&$E_{ls}$&$\beta_2$&$Q_{20}$&\multicolumn{3}{c||}
{Cluster Analysis}&\multicolumn{2}{c|}{$\alpha$-GCM}\\
&[MeV]&&[fm${}^2$]&${\cal O}$[\%]&$d [fm]$&$\sigma$ [fm]&${\cal O}$[\%]
&$\sigma$ [fm]\\
\hline
SkI3&2.5&0.677&46.8&82&2.70&1.68&98&1.65\\
SkI4&2.7&0.669&46.2&81&2.64&1.69&97&1.64\\
Sly6&2.4&0.666&48.0&81&2.68&1.73&97&1.68\\
SkM*&4.6&0.593&38.0&71&2.20&1.68&82&1.62\\
NL3 &   &0.679&40.7&85&2.52&1.56&  &    \\
$\chi_m$&&0.671&41.8&68&2.60&1.60& &    \\
\hline
\end{tabular}
\end{table}

This nucleus also gave a good opportunity to test the overlap with $\alpha$-GCM cluster wave functions
as described in section~\ref{gcmmodel}. We first used the distance $d$ as a collective coordinate,
distributing 10 points in the interval between $d=1\,$fm and $d=5\,$fm and varying $\sigma$ to
maximize the overlap. The result is given in the ``$\alpha$-GCM'' column in the tables. The overlap
increases noticeably. This calculation was performed only for the Skyrme-forces.

It is remarkable that the optimum $\sigma$ values are smaller for the ``$\alpha$-GCM'' case. It may be
that for the single-distance analysis the $\alpha$-particles have to widen a bit to simulate the
effect of orbital motion; one should not, however, read too much into this as the difference in
overlap between the different $\sigma$ values is of the order of 3\%.

The AMD wave function also has a quite large overlap of 96\% with the collective
two-$\alpha$-cluster model space. The overlap between the AMD and the Mean--Field wave function is
a little bit smaller at 85\%.

\subsection{The Nucleus \carbon}

The surprising feature in this nucleus is the discrepancy in ground-state properties. While three of
the Skyrme forces used have a spherical ground state, SkI3 produces a strongly deformed oblate
shape. In order to understand this, a constrained axially symmetric calculation was also performed,
which showed that the structure is really quite similar in all cases: the potential is very flat as
a function of deformation, and it just happens that the oblate point is about half an MeV lower than
the spherical one for SkI3, while this is reversed for the other three forces.

\begin{table}[htb]
\caption[]{Physical properties of \carbon\ for the four Skyrme forces considered. The other
columns give the maximum overlap with an $\alpha$-particle configuration for a fixed triangle with
side length 0.01~fm.
\label{tab:C}}
\begin{tabular}{|c|c|c|c||c|c|c|}
\hline
Force&$E_{ls}$&$\beta_2$&$Q_{20}$&\multicolumn{2}{c|}{Cluster Analysis}\\
&[MeV]&&[fm${}^2$]&${\cal O}$[\%]&$\sigma$\\
\hline
SkI3&15.8&-0.256&-24.7&28&1.72\\
SkI4&21.6&0.000&0.00&1&1.66\\
Sly6&19.8&0.000&0.00&1&1.68\\
SkM*&23.6&0.000&0.00&1&1.64\\
NL3 &    &0.000&0.00&1&1.48\\
$\chi_m$&&0.129&11.4&2&1.64\\
\hline
\end{tabular}
\end{table}

The search for the optimal $n\alpha$-cluster configuration in all cases yielded the $\alpha$-particles very
close to the center of mass of the nucleus in a roughly triangular configuration. The overlap is
much larger for the oblately deformed nucleus, since there is a preferred spatial alignment for the
triangle in this case. A calculation with fixed triangular geometry yielded very similar results.

An interpretation with an $\alpha$-GCM cluster configuration was attempted in two ways. First, a number
of different sizes of the triangles were used. The configuration which gave converged results was 10
points distributed between 0.01~fm and 3~fm. This improved the overlap very little, for example to
30\% for SkI3.

Alternatively rotated configurations were included. Adding the Slater determinants for the triangle
rotated in its plane by 2--10 equally spaced angles changed the overlap by less than 1\%, adding
three-dimensional rotations also produced similarly small effects.

The overall disappointingly small overlap with $\alpha$-clusters in this nucleus appears to be due
to the large spin-orbit energy, with only the deformed state for SkI3 a better candidate for
clustering. Because the oblate and spherical states are close in energy and not separated by a
barrier, including vibrational correlations in the mean-field model should lead to an intermediate
value for the overlap that is quite comparable for all forces.

\subsection{The Nucleus \oxygen}
For this nucleus all forces produce a spherical ground state, so that in this case the mean-square
radius is also an interesting quantity to compare. It is shown in Table~\ref{tab:O} and it is
remarkable that the radii agree very well for all four forces.

The cluster analysis with free placement produced similar results to the spherical \carbon\ nuclei:
the $\alpha$-particles tended to the same position at the center off mass, indicating that one is
not dealing with well-separated clusters but again with wave functions produced mainly by
antisymmetrization. The optimization located the particles at positions $(d,0,0)$, $(0,d,0)$,
$(0,0,d)$, and $(-d,0,-d)$ with $d=0.01$ (as mentioned, results are insensitive to the exact value
of $d$). This is not a natural arrangement, but seems to indicate that the principal requirement is
that the four particles not be coplanar. The small variation in the $\sigma$-values reflects that in
the mean-square radii.

\begin{table}
\caption[]{Properties of \oxygen\ for the four Skyrme forces
together with results for free cluster placement.\label{tab:O}}
\begin{tabular}{|c|c|c||c|c|}
\hline
Force&$E_{ls}$&$R_{\text rms}$&\multicolumn{2}{c|}{Cluster Analysis}\\
&[MeV]&[fm]&${\cal O}$[\%]&$\sigma$ [fm]\\
\hline
SkI3&1.0&2.65&96&1.76\\
SkI4&1.0&2.65&96&1.76\\
Sly6&0.9&2.69&96&1.79\\
SkM*&1.1&2.68&96&1.78\\
NL3 &   &2.56&95&1.72\\
$\chi_m$&&2.58&79&1.71\\
\hline
\end{tabular}
\end{table}

A cluster analysis was attempted using a tetrahedron of
$\alpha$-particles, with the particles' radii and mutual distance as
parameters. The results was that maximizing the overlap drives this
distance towards zero at $\sigma\approx1.76$fm. In this case the
limiting overlap is quite close to 1, showing that the tetrahedral
symmetry is a good basis for expanding the true wave functions.  Like
in the case of \carbon, however, the cluster interpretation is thus
invalid in this case. It should be noted that the overlap, which
reaches 98\% for a distance of 0.01~fm, only goes down to 97.5\% at
0.2~fm. This opens the alternative of either using \oxygen\ wave
functions or the four-$\alpha$ structure to look for the presence of
an \oxygen\ core in heavier nuclei.

That the precise arrangement of the clusters is not important was also
demonstrated by using fixed geometry. Using an exact tetrahedron
configuration produced identical results; only when the particles were
coplanar the overlap vanished immediately.

Our results are in complete agreement with the cluster model of \cite{zhang}.

In view of the good description of this nucleus by static cluster wave
functions we did not perform multiconfiguration studies in this case.

\begin{figure}[htb]
\includegraphics[width=4cm]{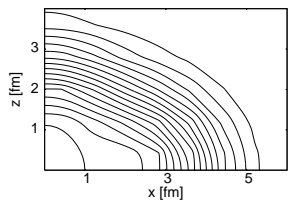}
\caption{Density contour lines for \neon.}\label{fig:Ne}
\end{figure}

\subsection{The Nucleus \neon}
In this case the shape of the nucleus, shown in Fig.~\ref{fig:Ne} shows
a strong prolate deformation with a sugestion of a central, more or
less spherical part, with two additional mass distributions added on
both sides. The physical properties are given in Table~\ref{tab:Ne}.
This would suggest a \carbon\ core with two $\alpha$-particles
added, and this is indeed what was produced by the unconstrained
search. This is quite close to the interpretation in \cite{zhang}, although there the central
\carbon\ has an annular structure.

\begin{table}[htb]
\caption[]{Physical properties of \neon\ for the four Skyrme forces
considered. The parameters for the cluster analysis are described in
the text. \label{tab:Ne}}
\begin{tabular}{|c|c|c|c||c|c|c|c|}
\hline
Force&$E_{ls}$&$\beta_2$&$Q_{20}$&\multicolumn{4}{c|}
{Cluster Analysis}\\
&[MeV]&&[fm${}^2$]&${\cal O}[\%]$&$d$ [fm]&$\sigma_1$ [fm]
&$\sigma_2$ [fm]\\
\hline
SkI3&8.5&0.423&91.0&53&1.91&1.78&1.71\\
SkI4&9.2&0.412&88.0&49&1.86&1.78&1.71\\
Sly6&8.5&0.409&89.8&47&1.84&1.80&1.74\\
SkM*&11.1&0.371&79.2&36&1.59&1.77&1.73\\
NL3 &    &0.425&84.4&59&1.83&1.74&1.65\\
$\chi_m$&&0.439&90.6&50&1.89&1.74&1.69\\
\hline
\end{tabular}
\end{table}

It appeared expedient to try different radii for the $\alpha$-particles in this case. Letting the
program optimize all the radii simultaneously instead of using a common $\sigma$ led to an
improvement of $<2$\% in overlap and naturally showed the three particles in the core with the same
radius, while the two displaced ones also were identical. The final configuration, for which details
are listed in the table, thus has three $\alpha$-particles with radius $\sigma_1$ in a triangle at
the central position and perpendicular to the longest axis of the nucleus, while two with radius
$\sigma_2$ are positioned at $\pm d$ along that axis.

An attempt with a multiconfiguration analysis yielded disappointing results: rotating the central
triangle gave less than 1\% improvement, while allowing the outer particles to move along the axis
yielded up to 2\%. Adding different sizes of the central triangle gave similar values.

The same 5$\alpha$-configuration is also obtained in an AMD calculation without parity projection
before variation. The overlap with the pure $n\alpha$ configuration amounts to 57\%. A larger
overlap of 61\% is obtained by including parity projection before variation. In this case, the
optimum configuration corresponds to an $\alpha$+\oxygen\ cluster nature. The overlap between the
AMD and Mean--Field wave functions is 78\% and 37\% without and with parity projection,
respectively.

\subsection{The Nucleus \magnesium}
\begin{figure}[htb]
\includegraphics[width=4cm]{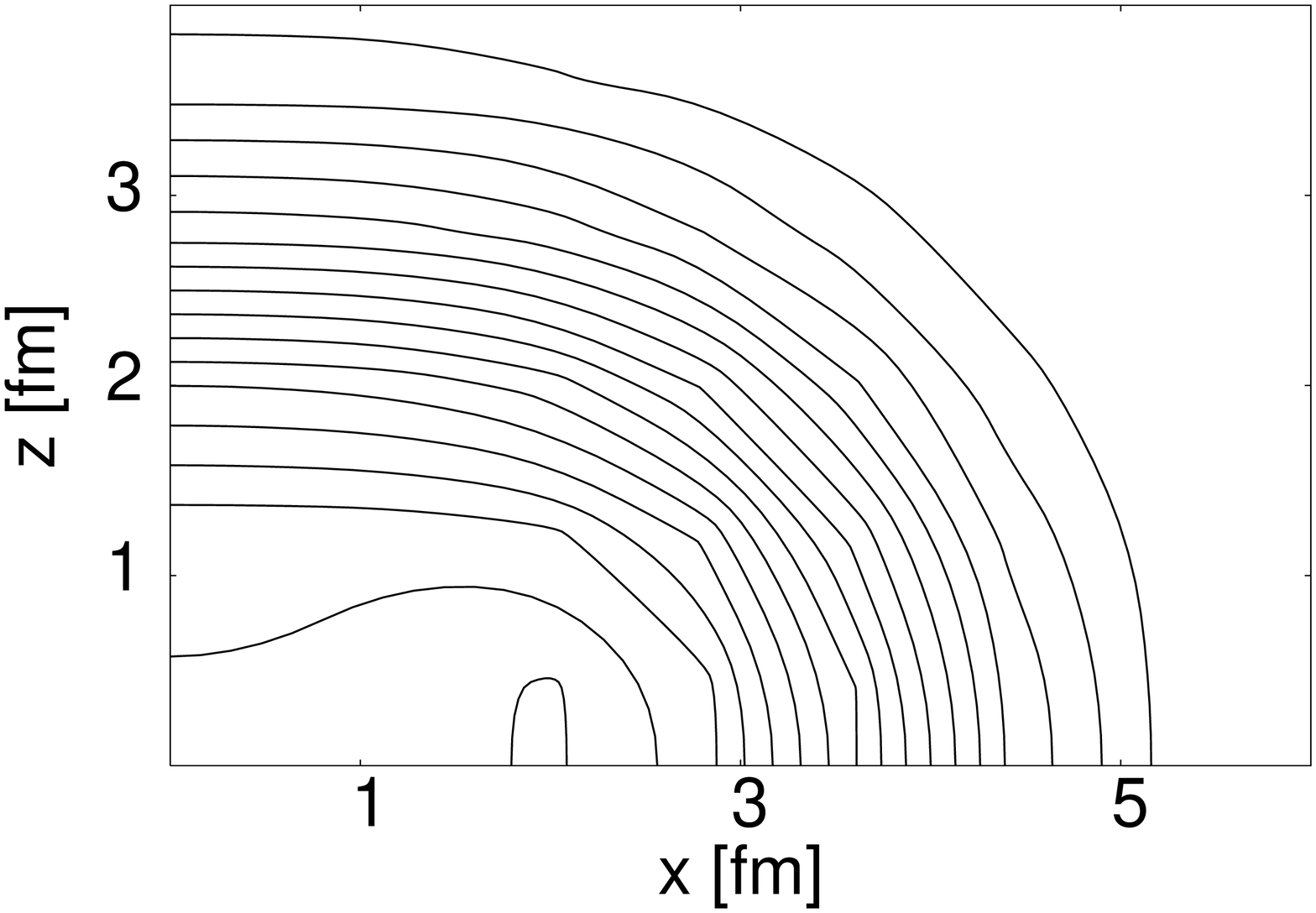}~~
\includegraphics[width=4cm]{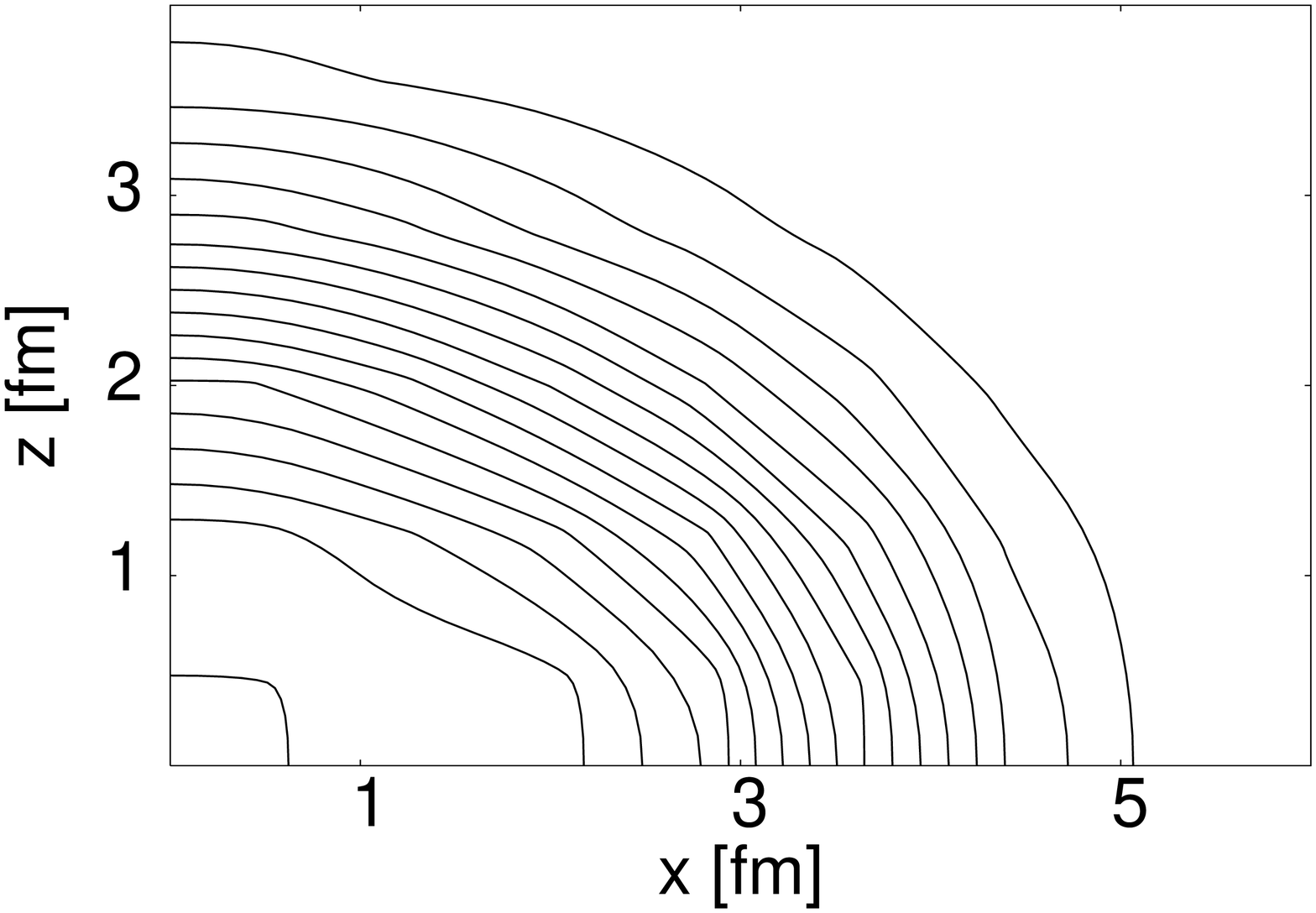}
\caption{Density contour lines for \magnesium\ (left) and \silicon (right).}\label{fig:MgSi}
\end{figure}

This nucleus is characterized by a strong quadrupole deformation for all forces considered, as seen
in Table~\ref{tab:Mg} and in Fig.~\ref{fig:MgSi}.

The unconstrained positioning of the $\alpha$-particles led to a grouping in doublets in the
following way: with the $x$-axis being the symmetry axis, two doublets are arranged at $x=\pm p$
with the particles displaced at slightly positive and negative $y$-values (around $\pm0.01$), while
the third doublet is obliquely shifted from the axis to $\pm(0,q,r)$ with $p$ and $q$ surprisingly
at larger values. While this appears to show some indications of true clustering, the overlap is
disappointingly small. The oblique positioning of the central doublet again indicates the need for
having the spatial degrees of freedom sufficiently represented.

\begin{table}[htb]
\caption[]{Physical properties of \magnesium. The parameters for the
cluster analysis are described in the text. \label{tab:Mg}}
\begin{tabular}{|c|c|c|c||c|c|c|c|c|}
\hline
Force&$E_{ls}$&$\beta_2$&$Q_{20}$&\multicolumn{5}{c|}{Cluster Analysis}\\
&[MeV]&&[fm${}^2$]&$\cal O$[\%]&$p$ [fm]&$q$ [fm]&$r$ [fm]&$\sigma$ [fm]\\
\hline
SkI3&22.6&0.423&117.2&2.3&1.40&0.17&0.37&1.77\\
SkI4&23.5&0.416&114.1&2.2&1.38&0.10&0.41&1.76\\
Sly6&21.6&0.413&116.5&2.2&1.39&0.04&0.40&1.79\\
SkM*&25.4&0.389&107.0&1.8&1.31&0.15&0.41&1.77\\
NL3 &    &0.416&103.9&2.0&1.37&0.17&0.45&1.71\\
$\chi_m$&&0.431&115.2&2.0&1.38&0.25&0.31&1.74\\
\hline
\end{tabular}
\end{table}

Alternative interpretations with fixed geometrical shape were also explored:
\begin{enumerate}
\item Two equilateral triangles displaced to $x=\pm d$ and either
  aligned the same way perpendicular to the symmetry axis or rotated
  by $30^\circ$ relative to each other. In both cases the maximum
  overlap that could be achieved by varying the triangle size and $d$
  was $<0.016$. Taking the overlap with two \carbon\ nuclei displaced
  along the symmetry axis yielded very small overlap ($<10^{-9}$).
\item Placing a square with vertices at $x=0$, $y,z=\pm q$ and two
  additional particles at $x=\pm p$ yielded microscopic overlaps of
  about 0.0002. Clearly this arrangement does not catch the symmetry
  of the single-particle space near the center. \label{squ}
\item Attempts of putting an \oxygen\ nucleus into the center, accompanied
by two $\alpha$-particles on both sides, lead to much smaller overlap.
\end{enumerate}

The interpretation of Ref.~\cite{zhang} in this case does not agree well; they favor the second of the
three choices given. The large value of the spin-orbit energy in our case seems to make a cluster
expansion less applicable.

The strong effect of the spin-orbit force to dissolve the cluster structure is also confirmed by the
AMD analysis. When the spin-orbit force is switched off, the AMD wave function has a
2$\alpha$+\oxygen\ cluster structure and large overlap (56\%) with the $n\alpha$-cluster wave
function, but a small overlap of a few percent with the Mean--Field wave function. When the
spin-orbit is included, on the other hand, the situation turns around: the overlap with the pure
$n\alpha$-cluster wave function is reduced to a few percent, while that with the Mean--Field one
goes up to 68\%.

\subsection{The Nucleus \silicon}
Except for SkM*, which shows a relatively pure quadrupole deformation, all forces have this nucleus
oblate deformed but with a strong hexadecupole contribution visible in the high-density contour
lines (see Fig:~\ref{fig:MgSi}). The oblate deformation is in agreement with other calculations
\cite{jaqaman,fink,patra,nazar1}.

As suggested by this geometrical shape, the optimal placement resulted in 5 $\alpha$-particles in a
pentagon perpendicular to the symmetry ($z$-) axis and two further particles at symmetric positions
on either side. The distances between the particles again tended towards zero, and because of that
the pentagon was not a regular polygon. Doing the calculation with a regular pentagon gave identical
results for the overlap. The Mean--Field wave functions thus seem to indicate a preference for the
pentagon symmetry, especially in view of the fact that the overlaps are quite large in this case.

\begin{table}[htb]
\caption[]{Physical properties of \silicon\ and optimum cluster
  parameters.
\label{tab:Si}}
\begin{tabular}{|c|c|c|c||c|c|}
\hline
Force&$E_{ls}$&$\beta_2$&$Q_{20}$&\multicolumn{2}{c|}{Cluster Analysis}\\
&[MeV]&&[fm${}^2$]&${\cal O}[\%]$&$\sigma$ [fm]\\
\hline
SkI3&30.2&-0.318&-108.5&7.8&1.86\\
SkI4&33.6&-0.294&-98.6&4.5&1.85\\
Sly6&31.7&-0.289&-98.8&4.0&1.87\\
SkM*&40.6&-0.224&-73.7&0.8&1.83\\
NL3 &    &-0.303&-93.5&6.2&1.80\\
$\chi_m$&&-0.351&-117.9&14.9&1.85\\
\hline
\end{tabular}
\end{table}

\begin{table}[htb]
\caption[]{Physical properties of \sulphur\ for the four Skyrme forces
considered. \label{tab:S}}
\begin{tabular}{|c|c|c|c||c|c|}
\hline
Force&$E_{ls}$&$\beta_2$&$Q_{20}$&\multicolumn{2}{c|}{Cluster Analysis}\\
&[MeV]&&[fm${}^2$]&$\cal O$[\%]&$\sigma$\\
\hline
SkI3&29.6&0.222&90.6&1.6&1.87\\
SkI4&37.0&0.147&58.7&0.3&1.86\\
Sly6&40.0&0.000&0.0&0.04&1.87\\
SkM*&44.7&0.000&0.0&0.03&1.86\\
NL3 &    &0.228&86.24&1.8&1.83\\
$\chi_m$&&0.253&101.1&2.2&1.85\\
\hline
\end{tabular}
\end{table}

\begin{figure}[htb]
\includegraphics[width=4cm]{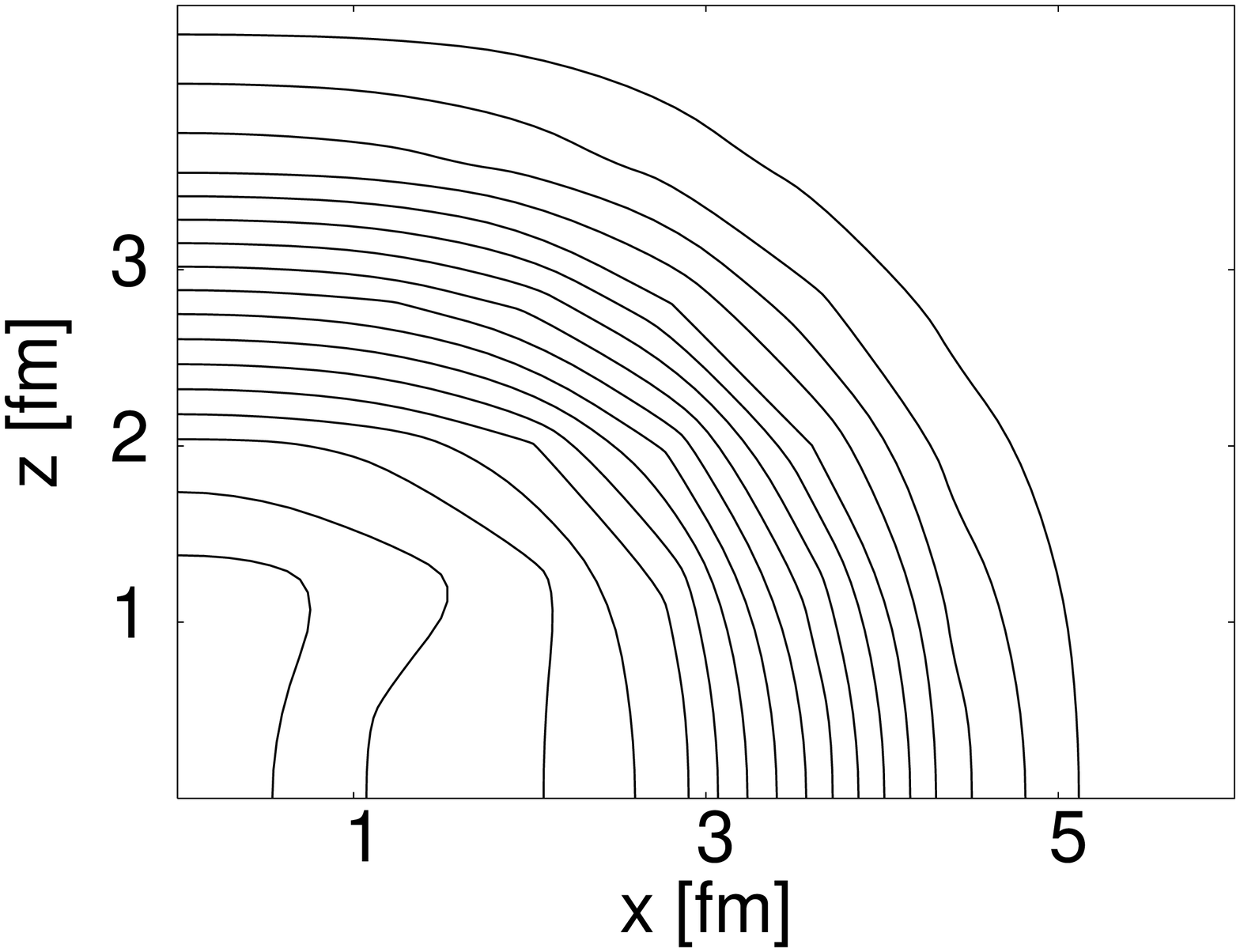}~~
\includegraphics[width=4cm]{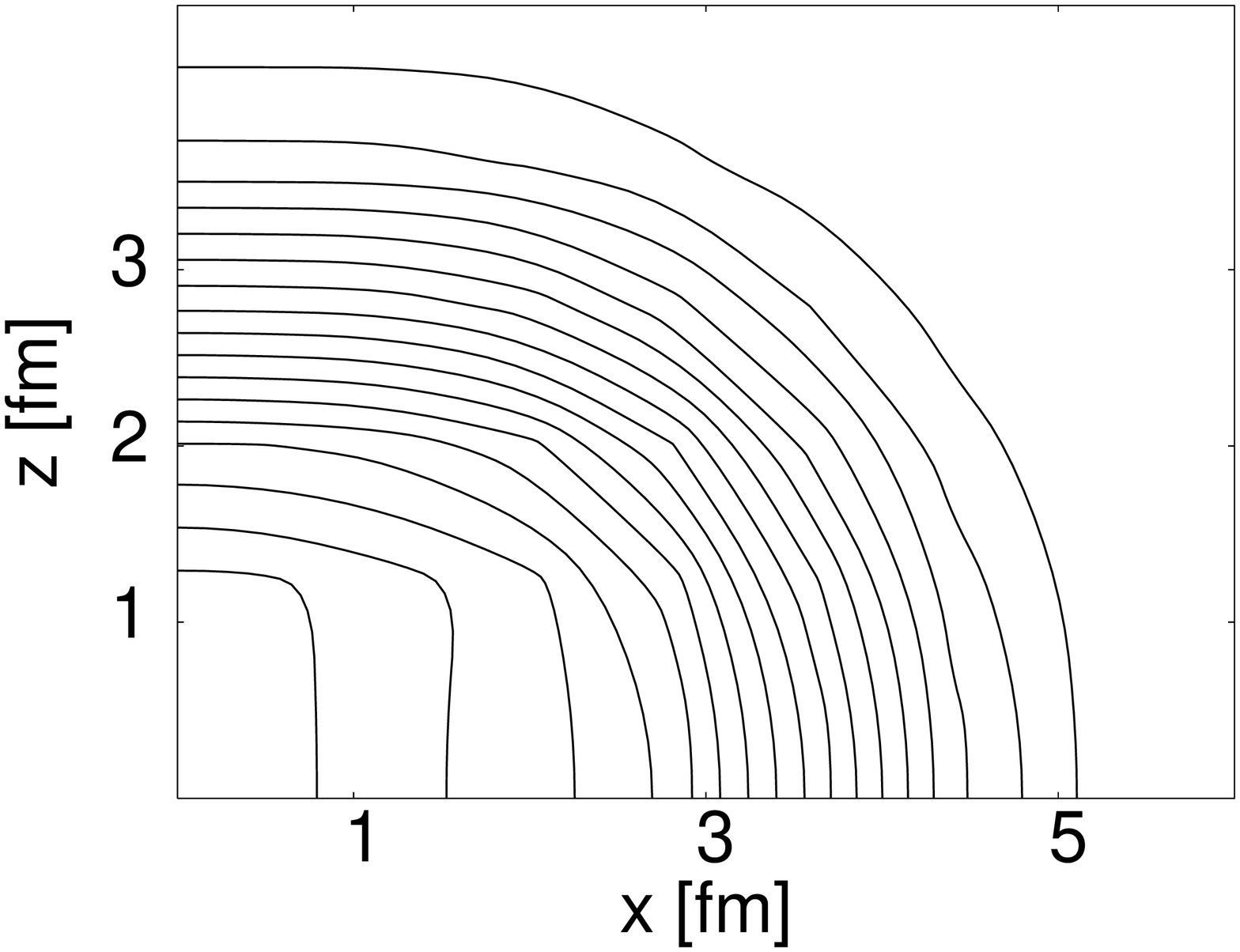}
\caption{Density contour lines for \sulphur\ and \argon.}\label{fig:SAr}
\end{figure}

\subsection{The Nucleus \sulphur}
This nucleus again shows a strong dependence on the force. For SkI3 and SkI4 it is prolate but with
the strange feature of an oblate central part at high densities (see Fig.~\ref{fig:SAr}), whereas it
is spherical for the other two forces. Other self-consistent calculations
(\cite{girod},\cite{jaqaman,fink,patra,nazar1}) generally prefer a prolate deformation.  The
unconstrained search for an $\alpha$-configuration in this case produced very small overlaps; the
geometric positions in all cases appeared almost random, filling a cube near the center of mass. The
distances, however, in this case were up to the order of 0.5~fm, showing that the overlap is quite
insensitive to the particle placement. The very small overlap may be due to the 1s$_{1/2}$ filling
and the large contribution of the spin-orbit force.

The AMD wave function in this case is closer to the Mean--Field one with an overlap of 40\%, but
has only a few percent overlap with the $n\alpha$-cluster wave function.

\subsection{The Nucleus \argon}
This nucleus appears as just the opposite of \sulphur: it is oblate with a deformation in agreement
with other calculations \cite{jaqaman,fink,patra,nazar1}, but shows an inner core of prolate
deformation at high density. The geometry of the $n\alpha$-cluster configurations found appeared as
random as in the case of \sulphur, but the distances to the center-of-mass were again of the the
order of $<0.1$~fm, which together with the much larger overlaps shows that is again is a case of a
dominating antisymmetrization effect. This comes close to the almost spherical geometry preferred by
\cite{zhang}.

It is, of course, somewhat unexpected that in this nucleus clustering appears stronger than for all
the others heavier than \neon. This should be due to the much reduced role of the spin-orbit
coupling in this nucleus as witnessed by the spin-orbit energy being only half of that for \sulphur.

\begin{table}[htb]
\caption[]{Physical properties of \argon\ and cluster analysis.
\label{tab:Ar}}
\begin{tabular}{|c|c|c|c||c|c|}
\hline
Force&$E_{ls}$&$\beta_2$&$Q_{20}$&\multicolumn{2}{c|}{Cluster Analysis}\\
&[MeV]&&[fm${}^2$]&$\cal O$[\%]&$\sigma$\\
\hline
SkI3&13.5&-0.183&-90.3&28&1.92\\
SkI4&15.2&-0.171&-83.8&21&1.91\\
Sly6&15.1&-0.162&-81.0&17&1.93\\
SkM*&19.1&-0.133&-65.9&7&1.92\\
NL3 &    &-0.186&-87.3&33&1.90\\
$\chi_m$&&-0.194&-94.1&29&1.89\\
\hline
\end{tabular}
\end{table}

\section{Strongly deformed states}\label{hyperdef}
\subsection{Technique for searching such states}
The search for strongly deformed isomeric states in the Mean--Field approximation has usually been
done with constraints. We tried a new method that has the potential for producing more exotic
configurations: initialization of the Mean--Field calculation with a fragment configuration. Thus,
e.~g., one may place two \oxygen\ nuclei at a certain distance $d$ from each other into a numerical
grid, orthogonalize, and then iterate to converge to a stationary state.  For small distances it may
converge into the ground state --- although we find this rarely happens, because the
antisymmetrization produces quite a different set of single-particle orbitals than that of the
ground state ---, for a large $d$ the nuclei may be driven apart, and finally, for some range
convergence may happen into a highly deformed configuration.

The attractive feature of this method is that initial configurations are highly flexible and might
lead into configurations that cannot easily be reached with a constraint. For example, a chain or a
triangle of three $\alpha$-particles may be used to look for a strongly deformed exotic state in \carbon,
or one might try various asymmetric mass configurations to to investigate whether they all converge
to the same deformed state.

With pairing omitted, we found a large number of exotic states, which, however, corresponded to
local minima under the constraint of a fixed single-particle configuration. Such minima may be
important for the dynamics, as they correspond to large collective mass when the collective motion
in the deformation is slow enough to cross over to another single-particle configuration, but
clearly are not isomeric states. We therefore used pairing in the iterations, which made most of the
minima disappear; the final states, however, as for the ground states, did not show significant
pairing.

\subsection{General features of superdeformed states}
A large number of different combinations of $n\alpha$-nuclei leading to compound systems up to
\calcium\ were considered and the static calculations with different initial fragment positions were
performed. Before going into the detailed results given in Table~\ref{tab:highdef} and the following
sections, let us summarize the main points of the findings:

\begin{itemize}
\item Superdeformed shapes were found in the majority of cases, but none of these appears to be of
clearly molecular type.
\item These exotic states were always axially and reflection-symmetric, even if the calculation was
started with two nuclei of different mass.
\item For the superdeformed states only Skyrme forces SkI3 and Sly6 were employed.  Generally the
two Skyrme forces yielded similar results, with the excitation energy somewhat more
force-dependent than the deformation. We only present results for SkI3 and mention discrepancies
to Sly6 where they appear interesting.
\item For these cases a comparison to the results of Ref.~\cite{zhang} is not given: although the
densities are roughly similar at least in the case of \sulphur, they did not use a spin-orbit
force, which  makes the comparison of the cluster interpretations problematic.
\end{itemize}

\begin{table}
\caption[]{Physical properties of highly deformed states for the
  Skyrme force SkI3 and the chiral model $\chi_m$.\label{tab:highdef}}
\begin{tabular}{|c|c|c|c|c||c|c|}
\hline
&$E^*$ & $E_{ls}$ &$\beta_2$& $Q_2$&\multicolumn{2}{c|}{Cluster Analysis} \\
System &[MeV] & [MeV] & & [fm$^2$]&${\cal O}$[\%]& $\sigma$ [fm] \\
\hline
\multicolumn{7}{|c|}{SKI3}\\
\hline
\silicon &13.2 & 25.6 & 0.773 & 325.9 & 1.3& 1.80\\
\sulphur &8.9 & 12.3 & 0.737 & 377.4 & 34.5 & 1.79\\
\argon & 8.7 & 29.1 & 0.529 & 283.4& 3& 1.88 \\
\calcium &26.3 & 28.1 & 0.983 & 859.8 & 0.8 & 1.80 \\
\hline
\hline
\multicolumn{7}{|c|}{$\chi_m$}\\
\hline
\silicon &11.1 & & 0.775 & 314.4 & 1.2& 1.77\\
\sulphur &5.7 &  & 0.743 & 368.6 & 36.2 & 1.75\\
\argon & 9.2 & & 0.533 & 278.9& 1.2& 1.84 \\
\calcium &23.7 & & 0.982 & 829.41 & 0.8 & 1.77 \\
\hline
\end{tabular}
\end{table}

The results are summarized in Table~\ref{tab:highdef}.

\subsection{The Compound System \magnesium}
In the combination of two \carbon\ nuclei, no indication of a superdeformed state was found. The
iterations always lead either to the ground state of \magnesium\ or to further separation.

\subsection{The Compound System \silicon}
\begin{figure}
\includegraphics[width=4cm]{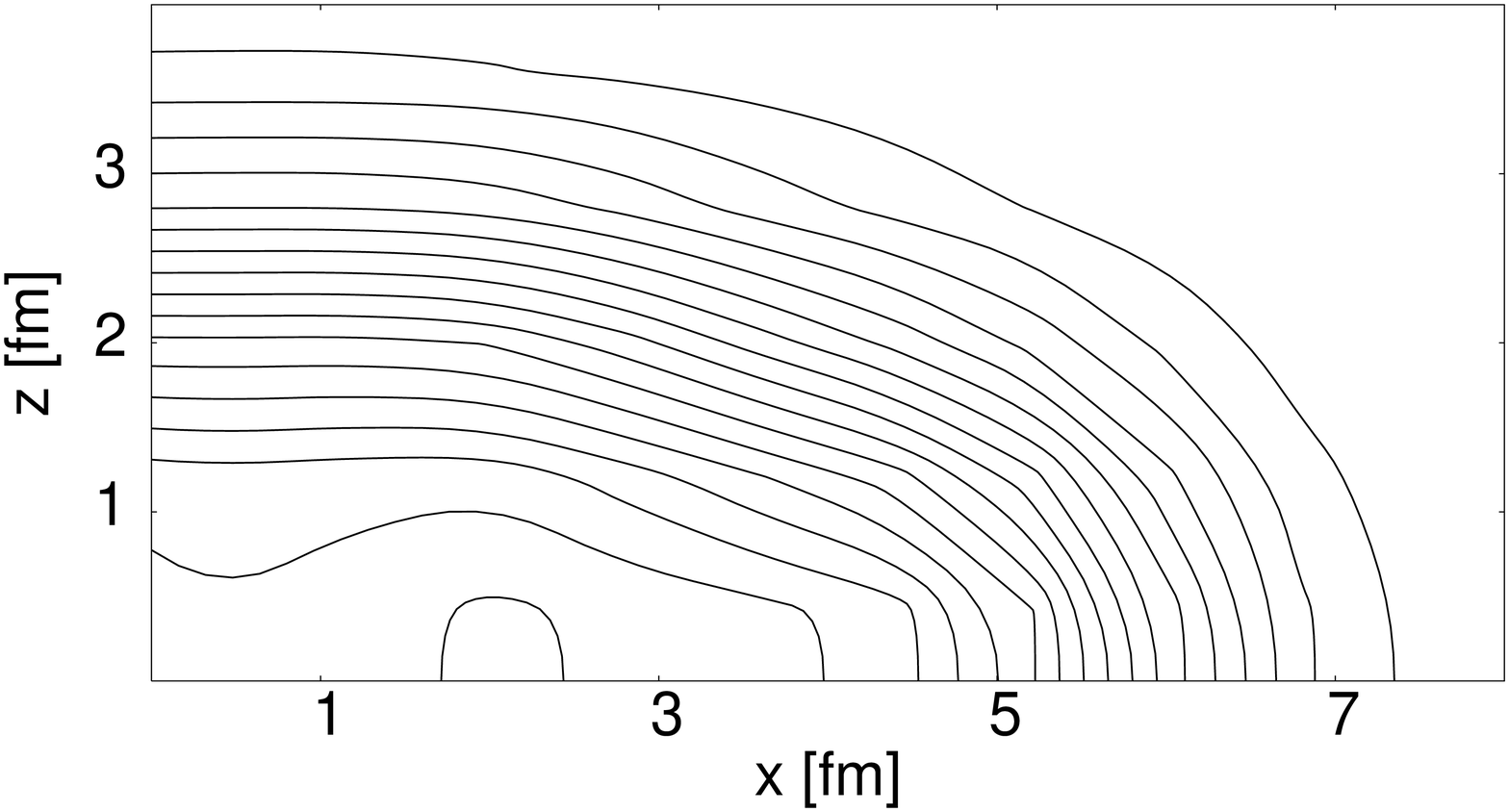}~~
\includegraphics[width=3.82cm]{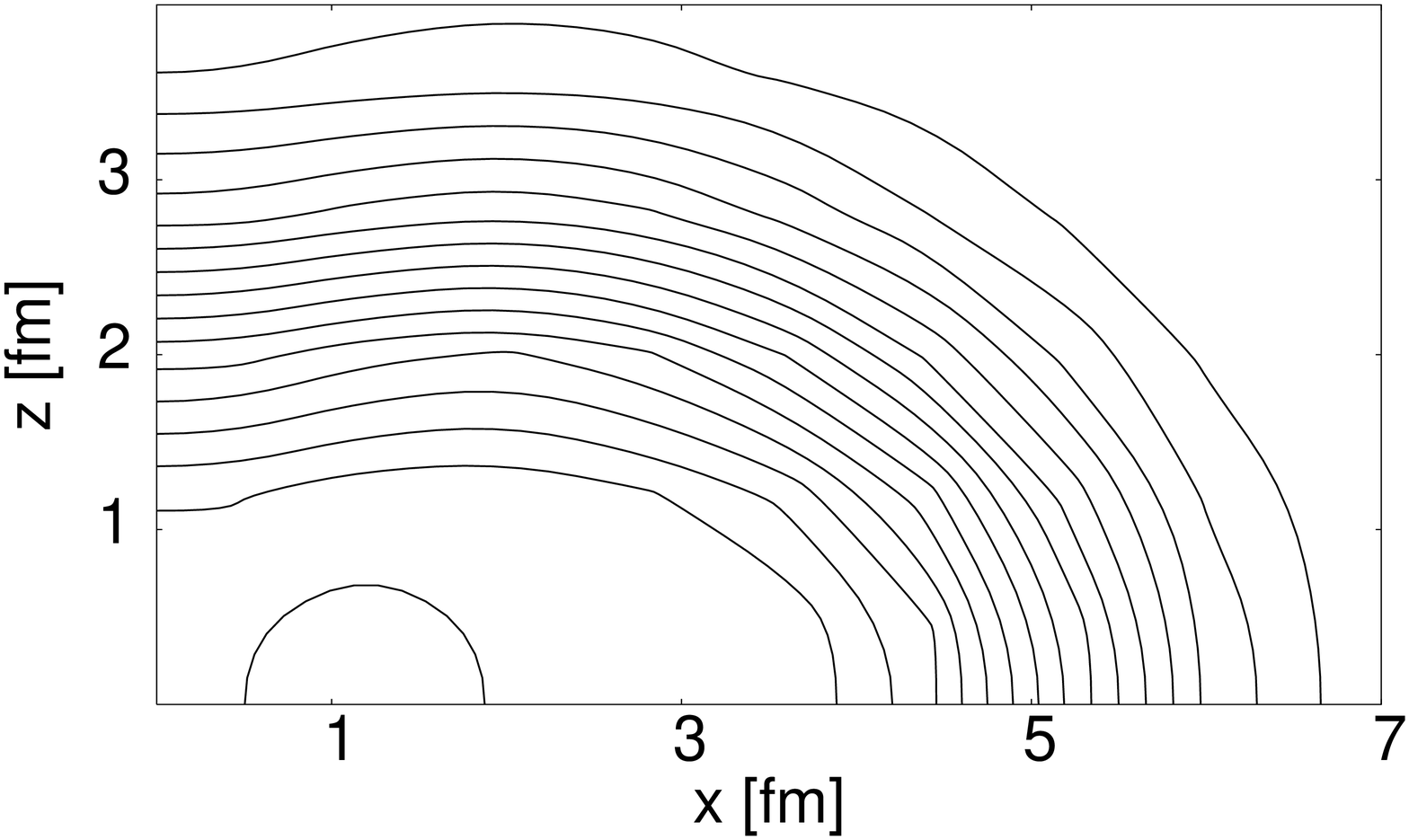}
\caption{Density contour lines for \silicon\ and \sulphur
superdeformed states.}\label{fig:SiSsuper}
\end{figure}

This is formed in the combination \carbon+\oxygen. The iterations lead -- aside from the ground
state of \silicon -- to one superdeformed state. Its properties, like of all the states found, are
summarized in Table~\ref{tab:highdef}. The density contour lines show a bit of a double-humped
structure at high density (see Fig.~\ref{fig:SiSsuper}). The unconstrained cluster analysis showed
that along the symmetry axis the particles are grouped in the following way: one is located at a
central $x$-position but with a half-fermi displacement in the $y$ and $-z$ directions (this
displacement is probably accidental and not significant for the overlap, since it violates overall
reflection symmetry). At $\pm1.23$~fm there are pairs of particles on both sides of the axis at a
relative distance of about 0.1~fm, while finally two particles are located at $\pm3.1$~fm on the
axis. The overlap with the $n\alpha$-cluster configuration is, however, quite small.

\subsection{The Compound System \sulphur}
This was studied in two different configurations: \carbon+\neon\ and \oxygen+\oxygen, so that an
interesting aspect is whether the initial asymmetry is lost. It was found that both configurations
converge rapidly into the same state, which is shown in Fig.~\ref{fig:SiSsuper}. A state of similar
deformation and excitation energy is well known in the literature with similar properties
\cite{flocard,doba1,yamagami,robledo,matsuyanagi}.  It shows a slight amount of necking-in,
making this the best candidate for a ``molecular'' configuration.

The relatively small spin-orbit energy, compared to the other superdeformed states, makes a cluster
analysis much more attractive.  The unconstrained $\alpha$-analysis shows all particles very close
(the usual 0.01~fm) to the symmetry axis with two triplets at $\pm1.9$~fm and additional single
particles at $\pm2.82$~fm. The center-of-mass of the four particles on each side is at $\pm2.1$~fm.
This is a clear indication of a molecular structure with somewhat distorted \oxygen\ nuclei at a
distance of about 4~fm.

\begin{figure}
\includegraphics[width=10cm]{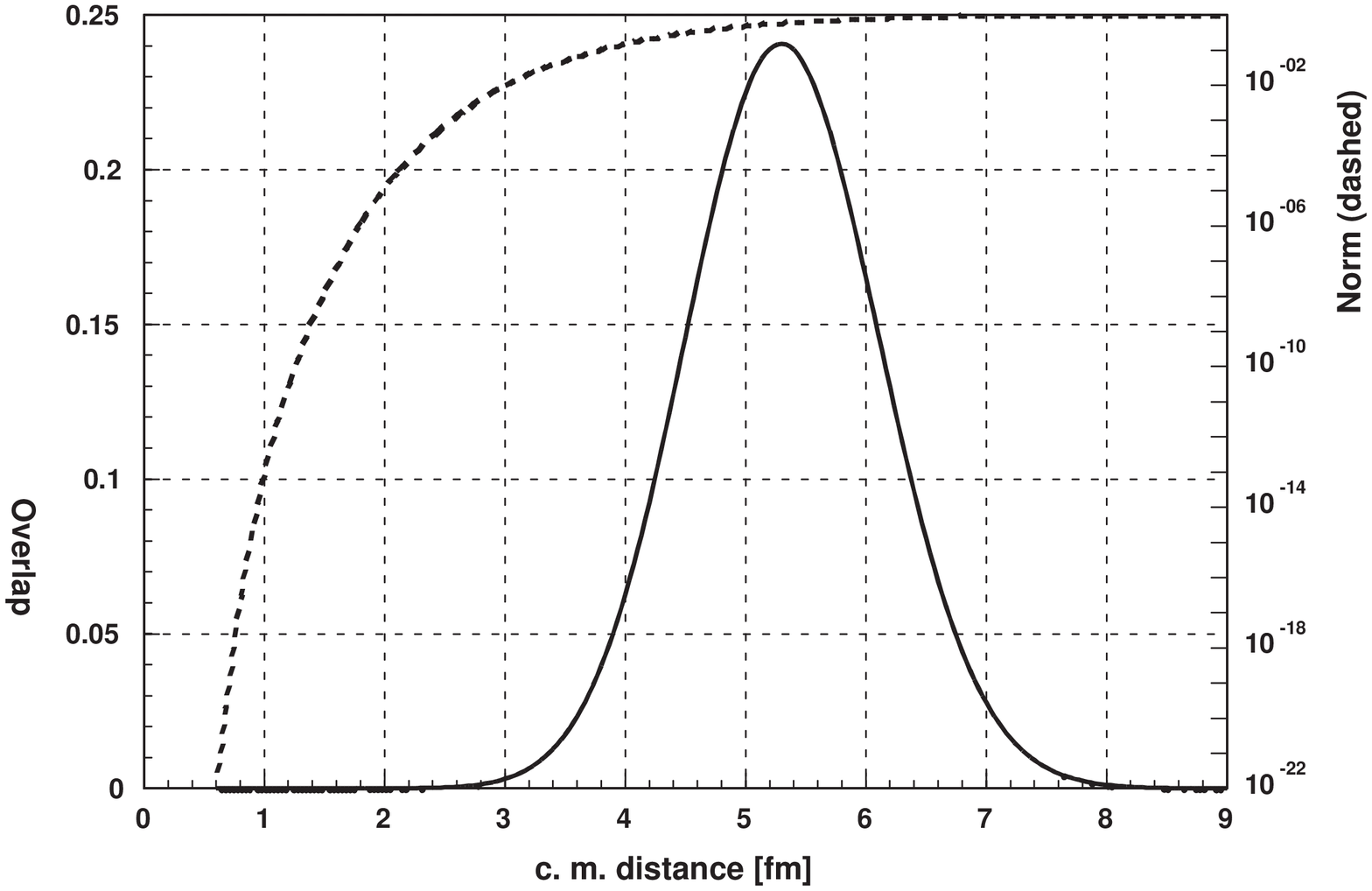}
\caption{Full curve: overlap of the strongly deformed state found in
  \sulphur\ with a configuration of two \oxygen\ nuclei placed symmetrically at a given
  distance. Dashed curve: norm of the combined Slater determinant of the two \oxygen\ nuclei.
  \label{fig:o16o16overlap}}
\end{figure}

For this reason an analysis of this system based on a double \oxygen\ configuration was also
attempted. Two ground-state configurations were inserted into the grid at a specified distance, with
the two nuclei being displaced symmetrically from the center-of-mass of the compound system along
the axis of symmetry. The resulting overlap is shown in Fig.~\ref{fig:o16o16overlap} together with
the norm of the combined Slater determinant. The resulting maximum overlap indicates that this
molecular configuration is not a perfect match; it is only near 6\% and, moreover, falls off
relatively slowly as the distance is varied.  On the other hand, the norm of the model wave function
(after orthogonalization) is already close to unity, indicating the possibility of molecular
structure at this distance, and using undistorted \oxygen\ nuclei for the overlap analysis is a
strong constraint.

The small overlap, however, is not the full story. Doing an analysis along the lines of
section~\ref{gcmmodel} with 20 different c.\ m.\ distances used between 3~and 8~fm, we find a total
overlap with this collective space of 35\% --- which agrees surprisingly well with the unconstrained
$\alpha$-particle analysis. The much more comprehensive analysis of this system done by Kimura and
Horiuchi \cite{kimura} gives a somewhat larger overlap of 57\%, but is based on a very different
wave function and technique, i.~e., a comparison of deformed-basis AMD \cite{defamd} with
generator-coordinate method to a resonating-group wave function for the double-\oxygen\ system. It
is not surprising that a mean-field description of this strongly deformed state contains less of an
$n\alpha$-cluster configuration than the AMD description. The AMD wave function, on the other hand,
was found to have a larger overlap with the mean-field one, also its spin-orbit energy is in
surprisingly good agreement with the mean-field one. This could be an interesting point for future
studies.

A similar attempt to increase the overlap with the $n\alpha$-cluster configuration by varying the
distance between the two quadruplet positions brought no further improvement.

\subsection{The Compound System \argon}
\begin{figure}
\includegraphics[width=4cm]{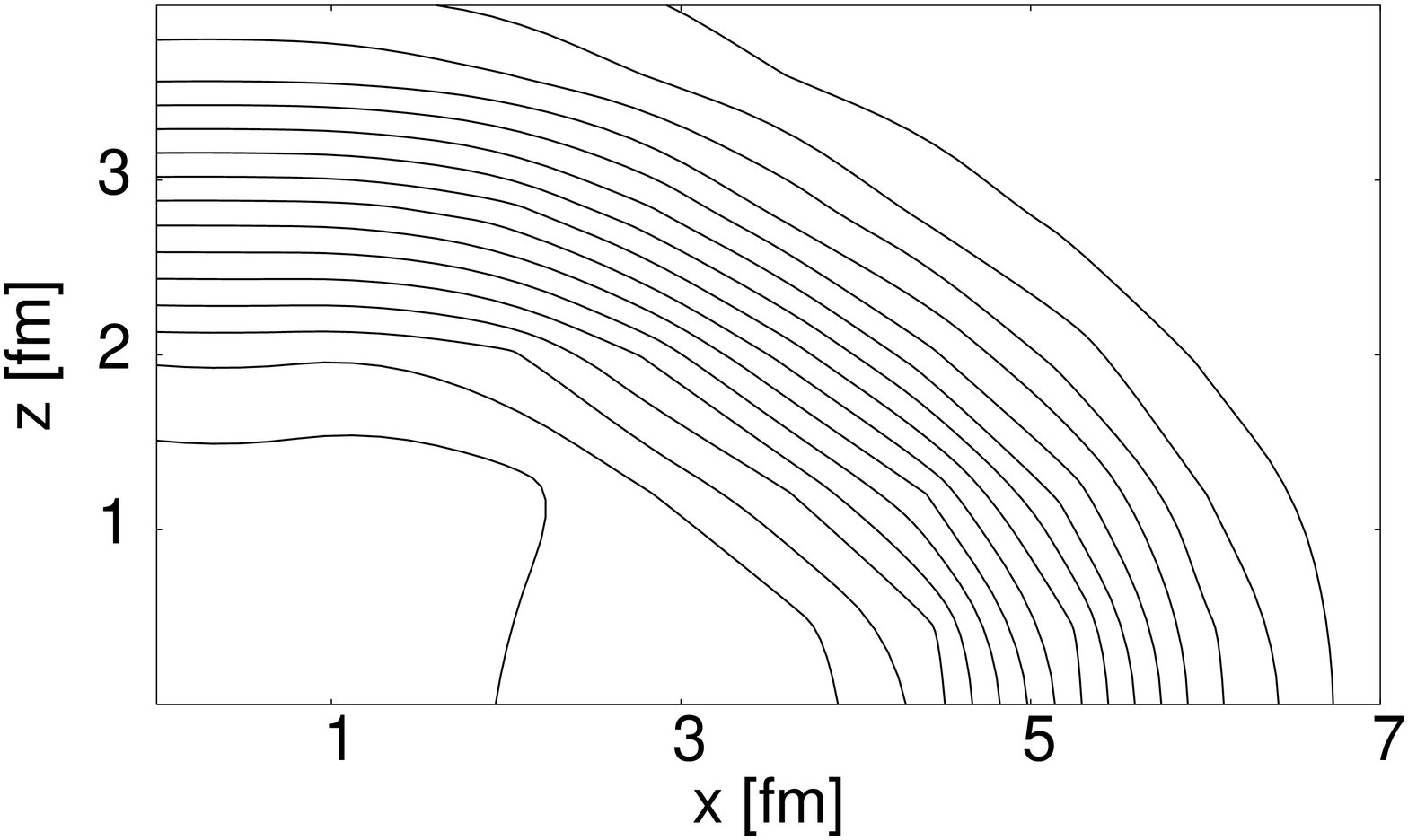}~~
\includegraphics[width=4cm]{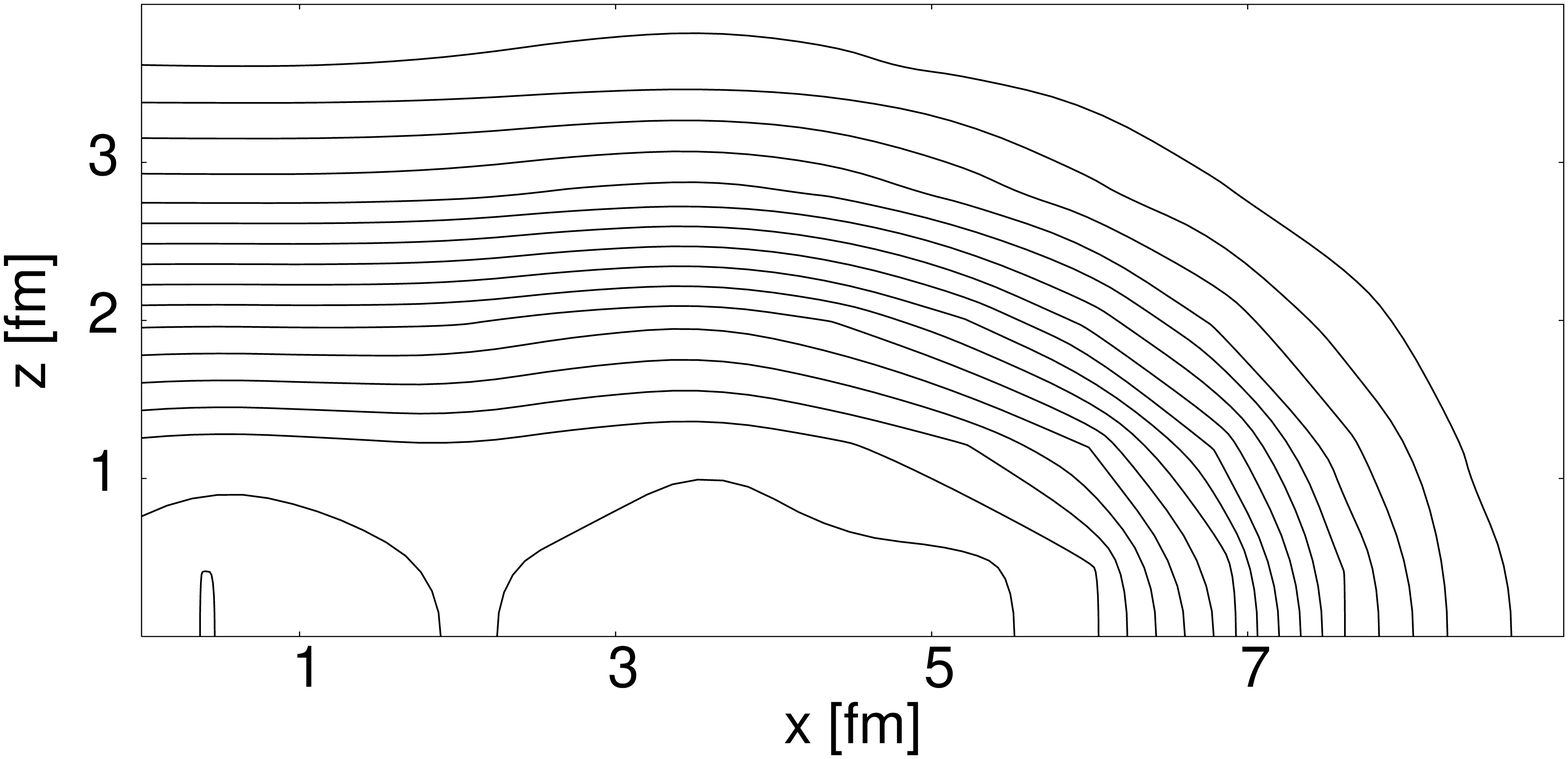}
\caption{Density contour lines for \argon\ (left) and \calcium\
(right) superdeformed states.}
\label{fig:ArCa}
\end{figure}

This is somewhat similar in showing a slight necking-in effect, and
again the iterations procedding from the two different initial
configurations \oxygen+\neon\ and \carbon+\magnesium\ led to
the same final state. The shape is shown in Fig.~\ref{fig:ArCa}. It is
interesting that for this system the calculation with pairing found
this superdeformed state rapidly from a wide range of initial
distances, whereas the unpaired calculation tended to get stuck near
$\beta=1$ and only relaxed to the correct state after several thousand
iterations.

The unconstrained $\alpha$-cluster analysis showed a diffuse
picture. The configuration yielding the overlap quoted had two
particles at $\pm2.7$~fm, a quadruplet in a distorted rectangle at
-0.3~fm and a triplet at +0.4~fm. The overlap depending only weakly on
the precise positioning, this is also consistent, e.~g., with the central
seven particles placed in two triangles facing each other and a single
$\alpha$ between at the center-of-mass.

\subsection{The Compound System \calcium}
This is really hyperdeformed as seen in Fig.~\ref{fig:ArCa}.  The density shows a triple-humped
structure, which is also borne out by the unconstrained placement of $\alpha$-particles. It produced
a configuration characterized by two particles near $\pm4.3$~fm, two particles at the center of
mass, and two triangular triplets near $\pm2.9$~fm along the symmetry axis. The overlap was quite
small, however, and seemed to depend only weakly on the exact placement of the particles, so that
the individual particles were up to half a fermi from symmetric positions along the axis; the
lateral displacement was always $\leq0.2$~fm, though. A further indication that at this point the
cluster interpretation becomes quite fuzzy is that for Sly6, which showed a practically identical
density distribution, one of the $\alpha$'s from the triangles move closer to those on the end
positions.

\section{Exotic $\alpha$-particle structures}

It should be briefly mentioned that the present code also allows the search for more exotic
$\alpha$-clustering structures, such as chains and polygons, by initializing the static
Mean---Field calculation with such a geometric arrangement. No interesting
state was found for 3 and 4 $\alpha$-particles, however, and as an example we confirmed the results
of Itagaki and collaborators \cite{itagaki} that a three-$\alpha$ linear chain is unstable with
respect to triaxial deformations also for the mean field.

\section{Summary}
The results obtained in this paper give a large amount of information
on the relationship between cluster wave functions and ground-state
Mean-Field ones. Let us summarize the main features of the
ground-state analysis first:
\begin{enumerate}
\item The automatic search for a $n\alpha$-cluster configurations with maximal overlap to the
  Mean-Field state worked surprisingly well and appears to really produce the best configuration,
  as all analyses with specific geometric assumptions showed. The geometry of the clusters appears
  similar to what is produced in cluster calculations.
\item Even when it was freely varied, the radius parameter for the $\alpha$'s always has reasonable
  values. It slowly increases with the mass of the nucleus, indicating the diminishing localization
  of the particles.
\item In all cases, the overlap decreased as the spin-orbit energy produced by a certain force
  parametrization increased. As expected the spin-orbit coupling destroys the spin-symmetric cluster
  structures.
\item The overlap with clusters also decreased when going to heavier systems. The growing effects of
  spin-orbit coupling and Coulomb explain that naturally.
\item Also as expected the cluster wave functions worked better for the lighter systems up to \neon.
  There was, however, again enhanced overlap for \silicon\ and \argon. It is interesting that the
AMD model deviates from a pure $n\alpha$-structure in the same way and stays closer to the mean-field
model.
\item In many cases the optimal $n\alpha$-structure contains groups of $\alpha$-particles spaced
very close together. They do not represent a true $\alpha$-cluster configuration, but correspond to
heavier clusters.
\item The attempts at including correlations in the form of collective superpositions were disappointing,
  producing only minor improvements in the overlaps.  \label{coll} Apparently their relatively large
  size compared to the total nucleus enables the clusters to fill the nucleus without additional
  spreading.
\item The relativistic model results were generally very similar to the Skyrme-force ones. The only
remarkable difference is in the cluster radii, which were systematically smaller for the
relativistic version. 
\end{enumerate}

For the superdeformed states the clearest example of cluster structure is \sulphur, and it is
especially interesting that the overlap with the double-\oxygen\ system agrees very well with that
of the unconstrained $\alpha$-configuration. 

\begin{acknowledgments}
This work supported by BMBF under contracts No. 06 F 131 and 06 ER 124.  We
gratefully acknowledge support by the Frankfurt Center for Scientific
Computing. One of the authors (JM) is indebted to the Japanese Society for the 
Promotion of Science (JSPS) for support during a visit to Kyoto University, where this
research was initiated.
\end{acknowledgments}

\end{document}